\documentclass[prd,showpacs,amsmath,amssymb,11pt]{revtex4}
\usepackage{latexsym}
\usepackage{bm}

\def\HollowBox #1#2{{\dimen0=#1 \advance\dimen0 by -#2
       \dimen1=#1 \advance\dimen1 by #2
        \vrule height #1 depth #2 width #2
        \vrule height 0pt depth #2 width #1
        \llap{\vrule height #1 depth -\dimen0 width \dimen1} 
       \hskip -#2
       \vrule height #1 depth #2 width #2}}
\def\BOX{\HollowBox{.100in}{.010in}}

\begin{document}

\title{Another proof of the positive energy theorem in gravity}

\author{{\" O}zg{\" u}r Sar{\i}o\u{g}lu}  
\email{sarioglu@metu.edu.tr}
\affiliation{Department of Physics, Faculty of Arts and  Sciences,\\
             Middle East Technical University, 06531, Ankara, Turkey}

\author{Bayram Tekin}  
\email{btekin@metu.edu.tr}
\affiliation{Department of Physics, Faculty of Arts and  Sciences,\\
             Middle East Technical University, 06531, Ankara, Turkey}

\date{\today}

\begin{abstract}
We show that gravitational energy expression simplifies when a new set of
coordinates that satisfies a certain asymptotic gauge condition is used. 
Compared to the ADM formula, positivity of the energy is more transparent in 
this new construction. Our proof relies on the asymptotic symmetries
(specifically supertranslations) of the asymptotically flat spacetimes at
spatial infinity in four dimensions.
\end{abstract}


\maketitle

\section{Introduction \label{intro}}

The positivity of the total energy of an asymptotically flat spacetime is
of extreme importance for the stability of the ground state (namely
the flat Minkowski spacetime). It is well known that unlike
other non-gravitational classical theories, having a positive energy
density everywhere does not automatically guarantee a total positive
energy of the spacetime in gravity. In fact, there are examples of
negative energy spacetimes which are not asymptotically flat: These 
include the so called $AdS$ (anti-de-Sitter) soliton \cite{horowitz}, 
Eguchi-Hanson soliton \cite{clarkson} and the $AdS$-Taub-NUT solitons 
\cite{stelea, hob}. In asymptotically flat spacetimes, the negative 
gravitational potential energy is not expected to dominate the initial 
``rest'' mass, {\em e.g.} crudely speaking, the negative gravitational 
potential energy of the Sun is about a million times smaller than its rest mass.

The positivity of the gravitational energy for asymptotically flat spacetimes
in \emph{four} dimensions was shown satisfactorily first by Schoen and Yau 
\cite{schoen} and later by Witten \cite{witten}. Schoen and Yau use 
complicated geometrical tools whereas Witten employs spinors and is 
inspired by supergravity for which the Hamiltonian is the square of a 
supercharge \cite{deser, grisaru}. We should note that Parker and Taubes 
\cite{parker} strengthened Witten's proof with more rigorous mathematical 
arguments, whereas Nester \cite{nester} corrected a technical error 
in Witten's calculation. In all of these works, the total energy that 
one starts with is the celebrated Arnowitt-Deser-Misner (ADM) \cite{adm} energy:
\begin{equation}
M_{ADM} = \frac{1}{16 \pi} \, \oint_{\Sigma} \, 
dS_{i} \, ( \partial_{j} h^{ij} - \partial^{i} \tilde{h} ) \, , \label{admass}
\end{equation}
where \( \tilde{h} = h^{j}\,_{j} \). The ADM formula is written in flat 
Cartesian coordinates, yet the final result, after the integration is 
carried out at spatial infinity, is gauge (or diffeomorphism) invariant.
[This statement, in fact, needs a major refinement: We have implicitly assumed 
a `proper' asymptotic behavior (or `appropriate' fall-off conditions on $h_{ij}$)
at spatial infinity such that the integral is well-defined. If this is not the case, 
then, even the Minkowski spacetime can have an arbitrary (positive, negative or even 
divergent) mass in some coordinate system. We show this in the appendix \ref{appa},
which somewhat generalizes the earlier work of \cite{den, bc, bartnik, chrus, mur}.] 
However, we stress that the integrand is \emph{not} gauge invariant. In these 
coordinates, the two terms in the integrand of (\ref{admass}) can be not only
of any sign but also compete with each other, and therefore one cannot 
see the positivity of energy by simply looking at (\ref{admass}); hence the 
complicated proofs mentioned above. We should note that the proofs of Schoen-Yau 
and Witten which employ the coordinate dependent integrand in the coordinate
independent ADM formula (\ref{admass}) work only in four dimensions and, to
our knowledge, no satisfactory generalizations have been given in 
generic dimensions.

The present work started with the hope of simplifying the proof of the positive 
gravitational energy theorem and extending it beyond \emph{four} dimensions. As 
will be seen shortly, even though simplification can be achieved by expressing 
the energy in a different gauge, a proof for generic $D$ dimensions is still elusive.

\section{\label{main} Choice of gauge}
We will start with a $D$-dimensional background diffeomorphism invariant expression 
for the integrand of the total energy and show that, as far as the positivity of 
the energy is concerned, the ADM expression is not a very convenient choice of gauge. 
In this gauge, one had to work hard to show that the negative energy 
contributions in (\ref{admass}) are dominated by the positive ones 
at the end. Analogous things happen in gauge theories: 
{\em e.g.} in QED, when one uses the Lorenz gauge, one has to
deal with propagating ghosts (namely negative norm states) and has to do
extra work (see {\em e.g.} the Gupta-Bleuler formalism \cite{gupta, bleuler}) 
to decouple these unphysical states. On the other hand, when one uses 
the Coulomb gauge, one works with only the physical states and the process
is devoid of such complications. Guided by this lesson from Quantum Field
Theory, we will look for an alternative to the Cartesian gauge of the
ADM expression which will surpass the difficulties faced in the proof
of the positive gravitational energy theorem. From the onset, it is
hardly clear as to what gauge there is to choose. However, we are lucky
that we already have the explicitly \emph{gauge invariant} total energy 
formula of \cite{db} which reduces to the ADM formula in the relevant limit
[See the discussion below for the details.]. We will show that with a 
proper choice of gauge (the coordinates), one ends up with a more transparent 
expression from which the positivity of the total energy for the 
$4$-dimensional asymptotically flat spacetimes should follow relatively 
easily.

Let us start by giving a brief outline of how gravitational charges are
defined. (We refer the reader to \cite{ad, db} for details. Even though
the discussion in these works is given for asymptotically $AdS$ spacetimes,
the formalism can safely be used for asymptotically flat spacetimes as well.) 
[For other constructions of charges in AdS spaces, see \cite{ht}.]
As usual, assume that the deviation, $h_{\mu\nu}$, of the actual 
$D$-dimensional static spacetime metric 
\( g_{\mu\nu} = \bar{g}_{\mu\nu} + h_{\mu\nu} \) from an 
asymptotically flat static metric (or the background) $\bar{g}_{\mu\nu}$
vanishes sufficiently rapidly at infinity. [We would like to stress that
for our purposes $\bar{g}_{\mu\nu}$ is flat in all senses but not
necessarily equivalent to $\eta_{\mu\nu}$.] One can then find a 
$D$-dimensional conserved and background gauge invariant charge 
(corresponding to each background Killing vector $\bar{\xi}^{\mu}$) 
which reads 
\begin{equation}
Q^{\mu} (\bar{\xi}) = \frac{1}{4 \Omega_{D-2}} \, 
\oint_{\Sigma} \, dS_{i} \, \sqrt{-\bar{g}} \, 
q^{\mu i} (\bar{\xi}) \, , \label{charge}
\end{equation}
where 
\begin{equation}
q^{\mu i} (\bar{\xi}) \equiv 
\bar{\xi}_{\nu} \bar{\nabla}^{\mu} h^{i \nu} -
\bar{\xi}_{\nu} \bar{\nabla}^{i} h^{\mu\nu} 
+ \bar{\xi}^{i} \bar{\nabla}_{\nu} h^{\mu\nu} 
- \bar{\xi}^{\mu} \bar{\nabla}_{\nu} h^{i \nu}
+ h^{\mu\nu} \bar{\nabla}^{i} \bar{\xi}_{\nu}
- h^{i \nu} \bar{\nabla}^{\mu} \bar{\xi}_{\nu}
+ \bar{\xi}^{\mu} \bar{\nabla}^{i} h
- \bar{\xi}^{i} \bar{\nabla}^{\mu} h
+ h \bar{\nabla}^{\mu} \bar{\xi}^{i} \, . 
\label{einchar} 
\end{equation}
This integral is to be evaluated at the $(D-2)$-dimensional boundary 
of a $(D-1)$-dimensional spatial hypersurface. The explicit gauge invariance
of (\ref{charge}) under time-independent gauge transformations, which is 
crucial for our arguments, is shown in \cite{db} [see the appendix \ref{appa}
for s subtle issue about `large' gauge transformations]. For a timelike 
Killing vector \( \bar{\xi}^{\mu} = (-1, {\bf 0}) \), 
$Q^0$ defines the total gravitational energy, which further reduces
to the ADM formula in flat Cartesian coordinates. For other types of
Killing vectors, one obtains different gravitational charges such as
the total angular momentum.

Since the background is assumed to be asymptotically flat and static, 
one can always choose the spatial boundary to be orthogonal to the time
direction and set $\bar{g}_{0i}=0$. We take the timelike Killing vector 
\( \bar{\xi}^{\mu} = (-1, {\bf 0}) \) and use it in (\ref{einchar}) by
making use of the identities
\begin{equation}
\bar{\nabla}_{j} \, \bar{\xi}_{0} = - \frac{1}{2} \partial_{j} 
\bar{g}_{00} \, , \qquad \qquad
\bar{\nabla}_{j} \, \bar{\xi}_{k} = 0 \, , 
\end{equation}
to obtain
\begin{equation}
E \equiv Q^0 = \frac{1}{4 \Omega_{D-2}} \, 
\oint_{\Sigma} \, dS_{i} \, \sqrt{-\bar{g}} \, 
\Big( \bar{\nabla}_{j} ( h^{ij} - \bar{g}^{ij} \tilde{h} ) 
- \frac{1}{2} ( h^{ij} - \bar{g}^{ij} \tilde{h} )
\bar{g}^{00} \partial_{j} \bar{g}_{00} \Big) \, . \label{ener}
\end{equation} 
In the ADM gauge, \( \bar{g}_{\mu\nu} = \eta_{\mu\nu} \),
the second term in (\ref{ener}) drops and the covariant
derivative in the first one turns into an ordinary derivative yielding
(\ref{admass}). However, at this stage we are free to choose other gauges, 
keeping, of course, $\bar{g}_{0i}=0$. Here we use this to
make the previously advertised gauge choice. Rather than being global, 
ours will only be an asymptotical one:
\begin{equation}
\bar{\nabla}_{j} ( h^{ij} + \bar{g}^{ij} \tilde{h} ) \sim O(1/r^{D-2+\epsilon}) \,  
\label{choi}
\end{equation}
for some $\epsilon > 0$ at the boundary $r \to \infty$. Note that the sign 
between the two terms is plus! Even though, 
\( \bar{\nabla}_{j} ( h^{ij} + \bar{g}^{ij} \tilde{h} ) = 0 \) holds exactly 
for the Schwarzschild solution presented in the Kerr-Schild coordinates as
\[ ds^2 = \eta_{\mu\nu} dx^{\mu} dx^{\nu} + \frac{2 m}{r} (dt - dr)^2 \, , \]
in general we only need (\ref{choi}) in what follows.

In fact, we were led to the gauge choice (\ref{choi}) by the
observation that the 4-dimensional Schwarzschild solution in its standard 
spherically symmetric form satisfies this gauge \emph{at the boundary}: 
\[ \bar{\nabla}_{j} ( h^{ij} + \bar{g}^{ij} \tilde{h} ) 
= \Big( - \frac{8 m^2}{r (r-2 m)^2}, 0, 0 \Big) \sim O(1/r^3) . \]
As for the Kerr solution written in the Boyer-Lindquist coordinates as
\begin{eqnarray*} 
ds^2 & = & -(1 - f(r,\theta)) \, dt^2 - 2 \, a \, f(r,\theta) \, 
\sin^{2}{\theta} \, dt \, d\varphi
+ \frac{n(r,\theta) \, dr^2}{h(r,\theta) \, (1 - f(r,\theta))} \\
& & + \frac{n(r,\theta) \, d\theta^2}{1 - f(r,\theta)} +
\Big( \frac{h(r,\theta) - a^2 \, \sin^{2}{\theta} \, f^{2}(r,\theta)}{1 - f(r,\theta)} \Big)
\, \sin^{2}{\theta} \, d\varphi^{2} \,, 
\end{eqnarray*}
where 
\[ f(r,\theta) = \frac{2 m r}{r^2 + a^2 \cos^{2}{\theta}} \,, \quad
   n(r,\theta) = r^2 - 2 m r + a^2 \cos^{2}{\theta} \,, \quad
   h(r,\theta) = r^2 - 2 m r + a^2 \,, \]
the background $\bar{g}_{\mu\nu}$ is found by taking $m=0$ and $a=0$, and one
finds after a rather long calculation that
\begin{eqnarray*}
\bar{\nabla}_{j} ( h^{ij} + \bar{g}^{ij} \tilde{h} ) & = &
\left( - \left( 66 a^{10} - 248 a^8 m r + 304 a^8 r^2 + 200 a^6 m^2 r^2 
- 872 a^6 m r^3 \right. \right. \\
& & + 64 a^4 m^3 r^3 + 558 a^6 r^4 + 192 a^4 m^2 r^4 - 1232 a^4 m r^5 + 512 a^2 m^3 r^5 \\
& & + 472 a^4 r^6 + 320 a^2 m^2 r^6 - 704 a^2 m r^7 + 112 a^2 r^8 + 256 m^2 r^8 \\
& & + a^2 \left[ 93 a^8-4 a^6 (93 m-95 r) r+64 a^2 r^4 (22 m^2 - 23 m r + 4 r^2) \right. \\
& & + a^4 r^2 (372 m^2 - 1396 m r + 523 r^2) \\
& & \left. + 16 r^5 (-32 m^3 + 60 m^2 r - 20 m r^2 + 5 r^3) \right] \cos{2 \theta} \\
& & + 2 a^4 \left[ 15 a^6+a^2 r^2 (92 m^2 - 140 m r + 33 r^2) + a^4 (-68 m r + 40 r^2) \right. \\
& & \left. + 4 r^3 (-8 m^3 + 24 m^2 r - 14 m r^2 + 5 r^3) \right] \cos{4 \theta} \\
& & \left. + \left[ 3 a^{10} - 12 a^8 m r + 4 a^8 r^2 + 12 a^6 m^2 r^2 - 12 a^6 m r^3 
+ 5 a^6 r^4 \right] \cos{6 \theta} \right) / \\
& & (8 r^3 ( a^2 + r (-2 m + r) )^2 (a^2 + 2 r^2 + a^2 \cos{2 \theta})^2 ) \, , \\
& & - a^2 \sin{2 \theta} \left( 15 a^6 - 54 a^4 m r + 61 a^4 r^2 + 48 a^2 m^2 r^2 
- 120 a^2 m r^3 + 96 a^2 r^4 \right. \\
& & + 32 m^2 r^4 - 96 m r^5 + 56 r^6 + a^4 (5 a^2 - 10 m r + 7 r^2) \cos{4 \theta} \\
& & \left. + 4 a^2 \left[ 5 a^4 + a^2 r (-8 m+17 r)  + 2 r^2 (-2 m^2 - 9 m r + 7 r^2) \right] 
\cos{2 \theta} \right) / \\
& & \left. \left( 4 r^4 ( a^2 + r (-2 m+r) ) (a^2 + 2 r^2 + a^2 \cos{2 \theta})^2 \right) \, , 0 
\right) \\
& \sim & O(1/r^3) \,,
\end{eqnarray*}
which shows that the gauge choice (\ref{choi}) is again satisfied.

Unfortunately though, our gauge is \emph{not} satisfied by the usual  
$D>4$ dimensional Kerr and Schwarzschild solutions: For example, in $D=5$
instead of the required $1/r^{3+\epsilon}$ behavior, these geometries 
have $1/r^3$ decays which renders our asymptotic gauge condition (\ref{choi}) 
inconvenient. Thus we will stick to four dimensions from now on. 

In addition to (\ref{choi}), one can also choose $\partial_{j} \bar{g}_{00} = 0$ 
for asymptotically flat spacetimes \footnote{At this point, one 
might mistakenly think that we have restricted the background to be just 
the flat Cartesian Minkowski spacetime, however this is not correct at all.}. 
Then the total energy expression (\ref{ener}) reduces to
\begin{equation}
E = - \frac{1}{8 \pi} \, 
\oint_{\Sigma} \, dS_{i} \, \sqrt{-\bar{g}} \, \partial^{i} \tilde{h} \, ,
\label{sener}
\end{equation} 
where this integral is to be evaluated on a large sphere $S^2$ at the spatial 
boundary yielding
\begin{equation}
E = - \frac{1}{8 \pi} \, 
\oint_{\Sigma} \, dS_{r} \, \sqrt{-\bar{g}} \, \partial^{r} \tilde{h} \, .
\label{ssener}
\end{equation}
To better appreciate (\ref{ssener}) instead of the full fletched gauge
invariant (\ref{einchar}), one can also directly compute the gauge-fixed
$00$ component of the linearized Einstein tensor $G_{L}^{00}$, from which 
the total energy follows 
\begin{equation}
\int d^{3} x \, T^{00} \, \bar{\xi}_{0} \, \sqrt{-\bar{g}} \sim
\int d^{3} x \, G_{L}^{00} \, \bar{\xi}_{0} \, \sqrt{-\bar{g}} \, .
\label{cav}
\end{equation}
The computation of $G_{00}^{L}$ is not so hard. We start by noting that
the assumptions made on the background so far yield a nontrivial 
$\bar{\Gamma}^{i}\,_{jk}$ and 
\( \bar{\Gamma}^{\alpha}\,_{\alpha j} = \partial_{j} (\ln{\sqrt{|\bar{g}|}}), \)
where all remaining background Christoffel symbol components vanish.
For the linearized Ricci tensor one finds
\[ R_{00}^{L} = \frac{1}{2} (- \bar{\BOX} \, h_{00} 
- \bar{\nabla}_{0} \bar{\nabla}_{0} \, h + 2 \bar{\nabla}^{\sigma} \bar{\nabla}_{0}
\, h_{\sigma 0} ) \, , \]
which upon explicit calculation simplifies to 
\[ R_{00}^{L} = - \frac{1}{2} \, \bar{g}^{ij} \, \bar{\nabla}_{i} \bar{\nabla}_{j}
\, h_{00} \, . \]
Following similar steps, one obtains
\[ R^{L} = - \bar{\BOX} \, h + \bar{\nabla}_{\alpha} \bar{\nabla}_{\beta} 
\, h^{\alpha \beta} \, , \]
where
\[ \bar{\nabla}_{\alpha} \bar{\nabla}_{\beta} \, h^{\alpha \beta} =
 - \frac{1}{\sqrt{-\bar{g}}} \, \partial_{i} (\sqrt{-\bar{g}} \, 
\bar{\nabla}_{j} \, h^{ij}) \qquad \mbox{and} \qquad 
\bar{\BOX} \, h = \bar{g}^{ij} \, \bar{\nabla}_{i} \bar{\nabla}_{j} \, h \, . \]
Thus
\[ G_{00}^{L} = R_{00}^{L} - \frac{1}{2} \, \bar{g}_{00} \, R^{L}
 = \frac{1}{2} \, \bar{g}_{00} \Big( \bar{g}^{ij} \, \bar{\nabla}_{i} \bar{\nabla}_{j} 
\, \tilde{h} + \frac{1}{\sqrt{-\bar{g}}} \, \partial_{i} (\sqrt{-\bar{g}} \, 
\bar{\nabla}_{j} \, h^{ij}) \Big) \, , \]
and employing our gauge choice (\ref{choi}) yields the suggestive form 
\footnote{There is a caveat here. To actually obtain (\ref{cav}), we have 
assumed that the gauge choice 
\( \bar{\nabla}_{j} ( h^{ij} + \bar{g}^{ij} \tilde{h} ) = 0 \) is valid 
everywhere in the bulk unlike ours which is only asymptotical. However
this does not change the final result since the neglected terms do not
contribute to the energy.}
\[ G_{00}^{L} = \bar{g}_{00} \, \frac{1}{\sqrt{-\bar{g}}} \, \partial_{i}
(\sqrt{-\bar{g}} \, \partial^{i} \, \tilde{h}) \sim T_{00} \,. \] 
Note that, unlike the ADM integrand which is made up of two competing terms with 
opposite signs, (\ref{sener}) is either positive or negative definite. 

By definition, \( \tilde{h} = \bar{g}^{ij} h_{ij} \) is the trace of a 
$3 \times 3$ positive definite matrix and thus is positive definite. This 
is so at least in the region where the original full metric $g_{\mu\nu}$ does 
not change its signature. For example, for a black hole $\tilde{h}$ could become 
negative definite inside the event horizon, however we are only interested in 
the behavior of $\tilde{h}$ near the spatial boundary. [See the discussion below 
as to what goes wrong when $\tilde{h}$ is negative definite \emph{everywhere}.] 
From the very assumption about the convergence properties of the deviation part 
of the metric, it follows that $\tilde{h} \to 0$ as $r \to \infty$. Moreover, 
being a positive definite quantity, it necessarily follows that its derivative at
the spatial boundary is negative which renders (\ref{sener}) positive; i.e.
\( \partial^{r} \tilde{h} < 0 \) and $E>0$. Thus the sign of $\tilde{h}$ 
at the spatial boundary determines the sign of the total energy. The 
crucial point in our argument is that, unlike the ADM case, our integrand 
does not change sign.

So what is wrong if $\tilde{h}$ is negative definite (in which case we
find $E<0$)? Recall that even for the Schwarzschild solution 
\( \tilde{h} \to - \tilde{h} \) also satisfies the (linear) Einstein 
equations. However, one sign is ruled out by the weak energy condition or 
by the assumption of the non-existence of naked singularities (see Witten 
\cite{witten}). This is also what happens in our case: If one choice of 
sign for $\tilde{h}$ leads to a negative energy, then this points out to 
the existence of a naked singularity and the sign of $\tilde{h}$ is to be 
reversed.

Going back to our examples, for the Schwarzschild solution one easily
finds $E=m$, when \( \tilde{h} = 2 m/(r- 2 m) \) is plugged into (\ref{ssener}).
As for the Kerr solution things are a little more complicated since
\[ \tilde{h} = \frac{a^2 \cos^{2}{\theta} }{r^2} 
+ \frac{2 m r - a^2 \sin^{2}{\theta} }{a^2 + r(r-2mr)} +
\frac{a^2 (a^2 + 2 m r + 2 r^2 + (a^2 - 2 m r) \cos{2 \theta})}{r^2 (a^2 + 2 r^2 + a^2 \cos{2 \theta})} \, . \]
However, substituting this into (\ref{ssener}) one correctly finds $E=m$ again.

The fact that the masses of the Kerr and Schwarzschild black holes can be
obtained using (\ref{ssener}) instead of the ADM expression is quite remarkable.
Moreover, the mass of any spacetime which asymptotically approaches either one 
of these two examples can be given by (\ref{ssener}) and will necessarily be 
positive as well. For more general spacetimes, we would now like to prove the 
existence of ``a gauge potential'' $\bar{\zeta}_{\mu}$ at the spatial boundary
$r \to \infty$ such that the new deviation part of the metric obtained by
\( h_{\mu\nu} \to h_{\mu\nu} + \bar{\nabla}_{\mu} \bar{\zeta}_{\nu}
+ \bar{\nabla}_{\nu} \bar{\zeta}_{\mu} \) satisfies the asymptotic gauge
condition (\ref{choi}) in \emph{four} dimensions, for which the energy
becomes (\ref{ssener}). It is well established that 
for asymptotically flat spacetimes the deviation part $h_{ij}$ should
decay like $ h_{ij} = O(1/r^{\alpha}) $ where $1/2 < \alpha$, otherwise
the energy definition becomes meaningless \cite{den, bc, bartnik, chrus}. 
So according to this, our gauge choice (in flat Cartesian coordinates)
\[ \partial_{j} h^{ij} + \partial^{i} \tilde{h} \sim O(1/r^{1+\alpha}) \, , \]
does not satisfy the $O(1/r^{2+\epsilon})$ requirement. However,
one can still employ a gauge transformation to take care of this. 
Under a gauge transformation, the left hand side of (\ref{choi}) (in 
flat Cartesian coordinates) implies that the gauge potential satisfies
\[ \partial^{2} \, \bar{\zeta}^{i} + 3 \, \partial^{i} \, \partial^{j} \, 
\bar{\zeta}_{j} \sim O(1/r^{1+\alpha}) \, , \]
which requires that $\bar{\zeta}^{i} \sim O(r^{1-\alpha})$. Such a gauge
potential $\bar{\zeta}^{i}$ always exists: In fact, suppose the right hand
side has the weakest decay given by the special form $c x^{i}/r^{2+\alpha}$.
Then 
\[ \bar{\zeta}^{i} = x^{i} \Big( c_1 - \frac{c_2}{r^3} 
 + \frac{c}{4} \frac{1}{\alpha(\alpha-3) r^{\alpha}} \Big) \, , \]
with $c_1$ and $c_2$ arbitrary real constants. Choosing $c_1 = 0$, we have
the required fall-off conditions. This is exactly what one needs to have an 
invariant mass definition under coordinate transformations at the asymptotically 
flat end of the spacetime. Such transformations are called ``supertranslations'' 
\cite{wald} and their required behavior was examined long time ago by the `asymptotic 
symmetries theorem' \cite{chrus}. This completes our proof that one can always choose 
(\ref{choi}) without changing the mass.

In conclusion, starting with a diffeomorphism invariant integrand for
the energy and choosing a more suitable gauge, we have given 
an alternative yet simpler total energy expression (\ref{ssener}) for
the 4-dimensional asymptotically flat static spacetimes, from which positivity 
of gravitational energy follows.

\appendix
\section{\label{appa} The mass of the Minkowski space}
Here we explain in detail how `large' coordinate transformations can give
mass to even a flat spacetime. We will write the standard Minkowski metric
in two different coordinate systems and compare one's mass relative to the
other one.

Consider the usual four dimensional Minkowski spacetime metric written in 
spherical coordinates \( ds^2 = -dt^2 + dr^2 + r^2 d \Omega_{2} \, . \)
Now make a coordinate transformation taking $t=f(\tau,\rho)$ and $r=h(\tau,\rho)$.
[A more restricted version of this was given beforehand in \cite{den, bc}.]
The metric becomes
\begin{equation} 
ds^2 = ({\dot{h}}^{2} - {\dot{f}}^{2}) \, d\tau^2 
       + 2 ( \dot{h} h^{\prime} - \dot{f} f^{\prime}) \, d\tau d\rho
       + ({h^{\prime}}^2 - {f^{\prime}}^2) \, d\rho^2 + h^2 d \Omega_{2} \, , 
\label{min}
\end{equation}
where prime and dot denote differentiation with respect to $\rho$ and $\tau$,
respectively. Needless to say, this is a flat metric with all the Riemann
tensor components vanishing. However does this metric have zero ADM mass?
The conventional version of the positive energy theorem assigns zero mass
uniquely to the flat Minkowski space. Here we will see that (\ref{min})
can have any energy. Setting $f=\tau$ and $h=\rho$ as the background and 
calculating the mass using (\ref{admass}) or (\ref{einchar}), one finds
\begin{equation} 
M = \frac{1}{2 \rho} [ -\rho^2 {f^{\prime}}^2 + (h- \rho h^{\prime})^2 ] \, . 
\end{equation}
In principle, depending on the choice of $f$ and $h$, $M$ can be any real
number. For example, when $f=\tau$, choosing $h = \rho H(\tau) + \kappa \sqrt{\rho}$,
with $H$ an arbitrary function of $\tau$ and $\kappa$ any real number, yields 
$M=\kappa^2/8$. In fact, if we choose a more general $h$ as 
$h=\rho + \kappa \rho^{1-s}$, $M$ will be divergent if $s<1/2$ and will vanish 
for $s>1/2$. Likewise, when $h=\rho$, choosing $f= F(\tau) + \kappa \sqrt{\rho}$, 
with $F$ an arbitrary function of $\tau$, one finds a negative $M=-\kappa^2/8$. 
In fact, one can examine the effect of such `large' gauge transformations on
the mass of any physically relevant solutions \cite{satek}. 

\begin{acknowledgments}
We thank S. Deser and M. G{\"u}rses for useful discussions and their 
critique on an earlier version of this manuscript. We would also like
to thank the referees for their useful remarks. B.T. thanks the Department 
of Mathematics at the UC Davis for hospitality, where part of this work
was conducted. 

This work is partially supported by the Scientific and Technological Research 
Council of Turkey (T{\"U}B\.{I}TAK). B.T. is also partially supported by the 
``Young Investigator Grant" of the Turkish Academy of Sciences (T{\"U}BA) 
and by the T{\"U}B\.{I}TAK Kariyer Grant No 104T177.
\end{acknowledgments}

\end{document}